# $q$-deformed Fourier Theory

JOCHEN SCHWENK*

*Max-Planck-Institut für Physik*
*Werner-Heisenberg-Institut*
*Föhringer Ring 6*
*D-80805 München, Germany*

## Abstract

We solve the problem of Fourier transformation for the one-dimensional $q$-deformed Heisenberg algebra. Starting from a matrix representation of this algebra we observe that momentum and position are unbounded operators in the Hilbert space. Therefore, in order to diagonalise the position operator in a momentum eigenbasis we have to study self-adjoint extensions of these operators. It turns out that there exist a whole family of such extensions for the position operator. This leads, correspondingly, to a one-parametric family of Fourier transformations. These transformations, which are related to continued fractions, are constructed in terms of $q$-deformed trigonometric functions. The entire family of the Fourier transformations turns out to be characterised by an elliptic function.

---

*E-Mail: jos@mppmu.mpg.de

# 1 Introduction

Based on non-commutative geometry the theory of quantum groups provides a generalisation of the symmetry principle in physics. A quantum group can be understood as a linear transformation acting on a space of non-commutative coordinates, usually called a quantum plane.

In order to investigate physical models with an underlying $q$-deformed symmetry one has to consider $q$-deformed Heisenberg algebras which can be derived from the differential calculus on quantum planes.

Then, as usual in quantum mechanical framework, the elements of these Heisenberg algebras have to be represented as self-adjoint operators in Hilbert space. By this procedure information of the possible eigenvalues of the dynamical variables is obtained.

In a first step, by using algebraic methods, one can construct matrix representations of the operators in question. This already shows that the spectra of position and momentum operators become discretised. The eigenvalues belong to a Fermat-spaced lattice, for example $p_{n+1} = qp_n$. This appears to be a common feature of $q$-deformed Heisenberg algebras [8, 3, 5].

Starting with a matrix representation in which the momentum operator is diagonal it is for many aspects important to find the representation in which the position operator is diagonalised. In this way the corresponding $q$-deformed Fourier transformation is defined.

In this paper this problem is solved for the one-dimensional $q$-deformed Heisenberg algebra. For this algebra a matrix representation on momentum eigenstates is known, which is recalled in section 2 of the paper. We find that the operators are symmetric and unbounded. Therfore, in order to diagonalise the position operator, self-adjoint extensions of it have to be investigated. Since for unbounded operators this is a non-trivial procedure, a brief account to this theory is given in section 3.

In section 4 we solve the eigenvalue equation of the position operator. The solutions, which can be characterised by continued fractions, show that there exists a one-parametric family of self-adjoint extensions. Each of these extensions corresponds to a different Fourier transformation. In sections 5 and 6 the entire family of these Fourier transformations is constructed in terms of $q$-deformed trigonometric functions. It turns out that the family of these transformations is characterised by an elliptic function, which allows to easily compute any scalar product in the position eigenbasis. By choosing a particular point on the torus we recover the Fourier transformation which is already known in the literature [4]. As a by-product we find addition theorems for the $q$-deformed trigonometric functions.

# 2 The $q$-deformed Heisenberg algebra

In [8] a $q$-deformed Heisenberg algebra has been introduced. It is generated by hermitian momentum and position variables $p$ and $\xi$ and an additional unitary element $u$, subject to the following algebraic commutation relations

$$\begin{aligned}
\frac{1}{\sqrt{q}}p\xi - \sqrt{q}\xi p &= -iu\,, \\
up &= qpu\,, \\
u\xi &= \frac{1}{q}\xi u\,, \\
\bar{p} = p\,, \quad \bar{\xi} = \xi\,, \quad u\bar{u} &= u\bar{u} = 1\,.
\end{aligned} \qquad (1)$$

Here $q > 1$ is the deformation parameter. These relations are based on the covariant differential calculus invented by Wess and Zumino [9]. The variable $u$ had to be introduced to allow hermitian $p$ and $\xi$. In the limit $q \to 1$ we have $u \to 1$ such that the usual algebra is recovered. A matrix representation of these operators on eigenvectors $\Phi_n$, $n \in \mathbb{Z}$ of $p$ is given in [8]

$$\begin{aligned} p\Phi_n &= p_n \Phi_n, \quad p_n = \pi_0 q^n, \\ u\Phi_n &= \Phi_{n-1}, \\ \xi\Phi_n &= \frac{-i}{p_n(q - \frac{1}{q})} \left( \frac{1}{\sqrt{q}} \Phi_{n+1} - \sqrt{q} \Phi_{n-1} \right). \end{aligned} \quad (2)$$

From the eigenvalues of $p$ it is clear that rescaling $\pi_0$ in the range $[\pi_0, q\pi_0)$ yields inequivalent representations. Introducing the scalar product $(\Phi_n, \Phi_k) = \delta_{nk}$ we get the Hilbert space

$$H_{\pi_0} = \left\{ \sum_{n=-\infty}^{\infty} c_n \Phi_n \; : \; \sum_{n=-\infty}^{\infty} |c_n|^2 < \infty \right\}. \quad (3)$$

Then by (2) the operators are defined on a vector space

$$D_{\pi_0} = \text{span}\{\Phi_n \; : \; n \in \mathbb{Z}\}, \quad (4)$$

which is a dense set in the Hilbert space $H_{\pi_0}$, i.e. $\overline{D_{\pi_0}} = H_{\pi_0}$. It is easily seen that $p$ and $\xi$ are symmetric operators and that $u$ is unitary, for example $(\xi\Phi_n, \Phi_k) = (\Phi_n, \xi\Phi_k)$ for all $n, k \in \mathbb{Z}$. From (2) it follows that $p$ and $\xi$ are unbounded symmetric operators in $H_{\pi_0}$, i.e.

$$\begin{aligned} \|p\Phi_n\| &= \pi_0 q^n, \\ \|\xi\Phi_n\| &\geq \frac{q^{-n}}{\pi_0(\sqrt{q} + 1/\sqrt{q})}, \end{aligned} \quad (5)$$

where $\|\Phi_n\| = 1$.

The problem of the Fourier transformation consists in finding an orthonormal basis of eigenvectors of the position operator $\xi$. One easily checks that in the domain $D_{\pi_0}$ there are no eigenvectors of the operator $\xi$. Therefore the operator $\xi$ has to be extended to a larger domain in which such a set of eigenvectors could be found. For unbounded, symmetric operators this is a non-trivial procedure since, in contrast to bounded operators, they cannot be extended to the entire Hilbert space. Under which conditions such extensions exist will be reviewed briefly in the next section. Additionally, by such an extension one gets a self-adjoint operator which has a unique spectral decomposition und is therefore interpretable as a physical observable.

## 3 Unbounded operators in Hilbert space

Let $T$ a, not necessarily bounded, linear operator with a dense domain $D_T$,

$$T : D_T \to H. \quad (6)$$

Then, an extension $T_1$ of $T$ is a linear operator $T_1 : D_{T_1} \to H$ with

$$D_T \subset D_{T_1} \quad \text{and} \quad T_1|_{D_T} = T, \quad (7)$$

which is usually described by $T \subset T_1$. We define the adjoint operator $T^*$ by considering all pairs $\{y, z\} \in H \times H$ with

$$(Tx, y) = (x, z) \quad \text{for all } x \in D_T. \quad (8)$$

This set $\{y, z\}$ is non-empty, it contains at least the pair $\{0, 0\}$. The set of all $y$ fulfilling (8) is called the adjoint domain $D_{T^*}$. From the fact that $D_T$ is dense in $H$ it follows that for each $y$ the corresponding $z$ is unique. Hence, one can define the *adjoint* operator $T^*$ by

$$T^* y = z \qquad \text{for } y \in D_{T^*}. \tag{9}$$

It is easily seen that for two operators $T \subset S$ it follows for the adjoints $S^* \subset T^*$.
A *symmetric* operator is defined by

$$(Tx, y) = (x, Ty) \qquad \text{for } x, y \in D_T, \tag{10}$$

from which we can conclude $T \subset T^*$ for symmetric $T$. If additionally

$$D_T = D_{T^*}, \tag{11}$$

then $T$ is called *self-adjoint*. A weaker condition for an operator $T$ is to be *essentially self-adjoint*, which means that its adjoint $T^*$ is self-adjoint, i.e. $T^{**} = T^*$.
If a symmetric operator is given, one can ask whether there exist self-adjoint extensions of it. Defining the deficiency indices $n_\pm$ of the operator $T$

$$n_\pm = \dim\{x \in D_{T^*} : T^* x = \pm \mathrm{i} x\}, \tag{12}$$

there is the following theorem

(i) $T$ is essentially self-adjoint if $n_+ = n_- = 0$.

(ii) $T$ has self-adjoint extensions if $n_+ = n_- \neq 0$.

(iii) $T$ has no self-adjoint extensions if $n_+ \neq n_-$.

In the case (ii) one can show that there is even a whole family of different self-adjoint extensions depending from as many parameters as the group $U(n_\pm)$. These extensions are in general characterised by different spectra of the operator. From this it is clear that for an operator to be interpretable as an observable it should be at least essentially self-adjoint. In this case there is the unique self-adjoint extension $T^*$, which has a unique spectral decomposition.

As an example we consider a symmetric operator $T$ with a dense domain $D_T$. We show that $T$ is essentially self-adoint if $D_T$ contains an orthormal basis $\{e_k\}$ of the Hilbert space $H$ on which $T$ is diagonal with real eigenvalues, $T e_k = a_k e_k$, $a_k \in \mathbb{R}$.
Since $\{e_k\}$ is a basis, any $y \in H$ has a decomposition $y = \sum y_n e_n$. We then consider (8)

$$(T e_k, \sum_n y_n e_n) = a_k y_k = (e_k, \sum_n a_n y_n e_n). \tag{13}$$

This must hold for all $k$ from which it follows $T^* \sum c_n e_n = \sum a_n c_n e_n$ for $\sum c_n e_n \in D_{T^*}$. Therefore $T^*$ is also a symmetric operator. Hence, we have $T^* \subset T^{**}$. We already know that $T \subset T^*$, because $T$ is symmetric. Applying the adjoint on the last equation we get $T^{**} \subset T^*$ and therefore $T^* = T^{**}$.
A more detailed account to the theory of unbounded operators can be found in standard textbooks [1, 7].

# 4 The eigenvalue problem

We study first the unbounded operator $p$ defined by (2) on the domain $D_{\pi_0}$. Since $p$ is diagonal on the basis $\{\Phi_n\}$ it is an essentially self-adjoint operator. For the domain of $p^*$ we find

$$D_{p^*} = \left\{\sum c_n \Phi_n \in H_{\pi_0} : \sum q^{2n}|c_n|^2 < \infty\right\}. \tag{14}$$

For the operator $\xi$ we have to examine the equation

$$(\xi \Phi_n, y) = (\Phi_n, z), \qquad \text{for all } n \in \mathbb{Z}, \tag{15}$$

where $z, y = \sum c_n \Phi_n \in H_{\pi_0}$. With help of (2) we then find for the adjoint operator

$$\xi^* \sum_{-\infty}^{\infty} c_n \Phi_n = \sum_{-\infty}^{\infty} \frac{-i}{q^n \pi_0 (q - \frac{1}{q})} \left(\frac{1}{\sqrt{q}} c_{n+1} - \sqrt{q} c_{n-1}\right) \Phi_n \tag{16}$$

on the domain

$$D_{\xi^*} = \left\{\sum c_n \Phi_n : \sum q^{-2n}|\frac{1}{\sqrt{q}} c_{n+1} - \sqrt{q} c_{n-1}|^2 < \infty\right\}. \tag{17}$$

In order to find self-adjoint extensions of $\xi$ we have to study the eigenvalue equation

$$\xi^* \sum_{n=-\infty}^{\infty} c_n \Phi_n = \lambda \sum_{n=-\infty}^{\infty} c_n \Phi_n, \tag{18}$$

wich leads by (16) to a recurrence equation

$$i\pi_0 \lambda (q - \frac{1}{q}) q^n = \frac{1}{\sqrt{q}} c_{n+1} - \sqrt{q} c_{n-1}. \tag{19}$$

For abbreviation we set further on $x_0 = \pi_0 \lambda (q - \frac{1}{q})$. The solutions of this equation are characterised by

**Theorem 1** *For each $x_0 \in \mathbb{C} \setminus \{0\}$ the recurrence relation*

$$ix_0 q^n c_n = \frac{1}{\sqrt{q}} c_{n+1} - \sqrt{q} c_{n-1} \tag{20}$$

*has exactly one convergent solution $\sum_{n=-\infty}^{\infty} |c_n|^2 < \infty$.*

*Proof.* This statement has already been proven in [2]. For completeness we give here a different proof.
Let $x_0 \in \mathbb{C} \setminus \{0\}$ arbitrarily fixed. First we show that for any solution of (20) the partial sum $\sum_{n=-\infty}^{0} |c_n|^2$ is finite.
Since $q > 1$ there exists $n_0$ such that

$$(1 + |x_0| q^n) < q^{\frac{1}{3}} \qquad \text{for all } n \leq n_0. \tag{21}$$

From the recurrence relation it follows for such $n \leq n_0$

$$\begin{aligned}
|\sqrt{q} c_{n-1}| &\leq |\frac{1}{\sqrt{q}} c_{n+1}| + |x_0| q^n |c_n| \\
&\leq (1 + |x_0| q^n) \max\{\frac{1}{\sqrt{q}} |c_{n+1}|, |c_n|\} \\
&\leq q^{\frac{1}{3}} \max\{\frac{1}{\sqrt{q}} |c_{n+1}|, |c_n|\},
\end{aligned} \tag{22}$$

and therfore $\max\{\sqrt{q}|c_{n-1}|, |c_n|\} \leq q^{-\frac{1}{6}} \max\{\sqrt{q}|c_n|, |c_{n+1}|\}$. A recursive use of this inequality for $n \leq n_0$ yields

$$\begin{aligned} |c_n| &\leq \max\{\sqrt{q}|c_{n-1}|, |c_n|\} \\ &\leq q^{-\frac{n_0-n}{6}} \max\{\sqrt{q}|c_{n_0-1}|, |c_{n_0}|\} = q^{\frac{n}{6}} M_0 \, , \end{aligned} \quad (23)$$

From which it follows that $\sum_{n=-\infty}^{n_0} |c_n|^2 \leq M_0^2 \sum_{n=-\infty}^{n_0} q^{\frac{n}{3}} < \infty$.

We are now going to construct a solution for which $\sum_0^\infty |c_n|^2 < \infty$. For this purpose we consider the following recurrence relation

$$\gamma_n + \frac{1}{\gamma_{n-1}} = x_0 q^n \, , \qquad \text{for } n > n_1 \, , \quad (24)$$

which actually follows from (20) by $\gamma_n = -\mathrm{i}\frac{c_{n+1}}{\sqrt{q}c_n}$. It is easy to check that a solution of this non-linear recursion is given by the continued fraction

$$\gamma_n = \frac{1}{x_0 q^{n+1}} \cfrac{1}{1 + \cfrac{a_{n+1}}{1 + \cfrac{a_{n+2}}{1 + \cfrac{a_{n+3}}{1 + \ddots}}}} \, , \quad (25)$$

where $a_n = -(x_0^2 q^n q^{n+1})^{-1}$. Such continued fractions are defined by the limit of their partial fraction, c.f. [10]. We have to show that $\gamma_n$ is well-defined, which means that the continued fraction (25) is converging. Since $q > 1$ we find a $n_1 \in \mathbb{N}$ such that

$$|a_n| = |x_0^2 q^n q^{n+1}|^{-1} \leq \frac{1}{4} \, , \qquad \text{for } n \geq n_1 \, . \quad (26)$$

Then, by Worpitzky's Theorem (see appendix), the continued fraction (25) is converging for all $n \geq n_1$ and its value is subject to the inequality

$$\left| (x_0 q^{n+1}) \gamma_n - \frac{4}{3} \right| \leq \frac{2}{3} \, , \quad (27)$$

from which it follows easily that

$$|\gamma_n| \leq \frac{2}{|x_0| q^{n+1}} \, , \qquad \text{for } n \geq n_1 \, . \quad (28)$$

Using these $\gamma_n$ we can build a solution for (20) by defining $c_n$ for $n \geq n_1$

$$\begin{aligned} c_{n_1} &= 1 \, , \\ c_{n+1} &= \mathrm{i}\sqrt{q}\gamma_n c_n \, , \qquad \text{for } n \geq n_1 \, . \end{aligned} \quad (29)$$

This solution to the linear recurrence relation (20) has the property

$$|c_n| = |(\mathrm{i}\sqrt{q})^{n-n_1} \gamma_{n-1} \gamma_{n-2} \cdots \gamma_{n_1}| \leq \left| \frac{2}{x_0} \right|^{n-n_1} q^{\frac{1}{2}(n_1^2 - n^2)} \, , \quad (30)$$

where we made use of (28). Because of the dominating factor $q^{-n^2}$ the sum $\sum_{n_1}^\infty |c_n|^2$ is convergent.

This particular series $c_n$ can be extended for all $n < n_1$ by applying the recursion (20) giving a solution with $\sum_{-\infty}^{\infty} |c_n|^2 < \infty$. The convergence at $n = -\infty$ is ensured by the first part of this proof.

We show finally that there is also a divergent solution to (20). Choose a $n_i \in \mathbb{Z}$ such that $\sqrt{q}|x_0|q^n < 1 + q$ for all $n \geq n_i$, and two initial values $d_{n_i}$ and $d_{n_i+1}$ with

$$|d_{n_i+1}| \geq q|d_{n_i}|, \tag{31}$$

which fixes all $d_n$, $n \in \mathbb{Z}$, because (20) is of second order. Let us assume that $|d_{n+1}| \geq q|d_n|$ for some $n > n_i$. Then it follows that also

$$|d_{n+2}| \geq \left| |\sqrt{q}x_0 q^{n+1} d_{n+1}| - |q d_n| \right| \geq \left(\sqrt{q}|x_0|q^{n+1} - 1\right)|d_{n+1}| \geq q|d_{n+1}|, \tag{32}$$

which shows by induction that $|d_{n+1}| \geq q|d_n|$ for all $n > n_i$. Therfore the sum $\sum |d_n|^2$ is divergent.
Since the linear recurrence relation (20) is of second order there is no other linear independant solution. $\square$

From the proof of this theorem we see that the convergence of a solution depends only from the ratio $c_N/c_{N+1}$ for some given $N \in \mathbb{Z}$, the other constant is an over-all normalisation.
In particular, we can conclude from this theorem that for the operator $\xi^*$ the dimensions of the eigenspaces to eigenvalue $\pm i$ are 1, respectively. Therefore $\xi$ is a symmetric operator with deficiency indices $n_+ = n_- = 1$, it has a one-parametric family of self-adjoint extensions.
We are now going to construct these extensions by finding all eigenvectors of $\xi^*$.

## 5 Eigenvectors of the position operator

We consider the following $q$-deformed trigonometric functions

$$\begin{aligned}
\cos_q z &= \sum_{n=0}^{\infty} \frac{(-1)^n z^{2n}}{[2n]!}, \\
\sin_q z &= \sum_{n=0}^{\infty} \frac{(-1)^n z^{2n+1}}{[2n+1]!},
\end{aligned} \tag{33}$$

where we use $[n]! = \prod_{k=1}^{n}\left(q^k - q^{-k}\right)$ and $[0]! = 1$. The functions $\cos_q z$ and $\sin_q z$ are continous in the variable $z$. In the limit $q \to 1$ they tend to the usual trigonometric functions in the sense $\cos_q\left((q - \frac{1}{q})z\right) \to \cos z$.
They are subject to the following $q$-difference equations

$$\begin{aligned}
\frac{1}{z}\left(\sin_q(qz) - \sin_q(\frac{1}{q}z)\right) &= \cos_q z, \\
\frac{1}{z}\left(\cos_q(qz) - \cos_q(\frac{1}{q}z)\right) &= -\sin_q z,
\end{aligned} \tag{34}$$

which can be verified directly from (33).
We make the following ansatz for an eigenvector $\Psi$ of the operator $\xi^*$

$$\begin{aligned}
\Psi &= \sum_{-\infty}^{\infty} q^n \left(A \sin_q q^{2n+a} + B \cos_q q^{2n+b}\right) \Phi_{2n} \\
&+ \sum_{-\infty}^{\infty} q^n \left(C \sin_q q^{2n+c} + D \cos_q q^{2n+d}\right) \Phi_{2n+1},
\end{aligned} \tag{35}$$

where $a, b, \ldots, C, D$ are arbitrary complex parameters. We will see later that this is a general ansatz. Substituting this in the eigenvalue equation

$$\xi^* \sum_{-\infty}^{\infty} c_n \Phi_n = \lambda \sum_{-\infty}^{\infty} c_n \Phi_n = \sum_{-\infty}^{\infty} \frac{-\mathrm{i}}{q^n \pi_0 (q - \frac{1}{q})} \left( \frac{1}{\sqrt{q}} c_{n+1} - \sqrt{q} c_{n-1} \right) \Phi, \tag{36}$$

and making use of (34) yields conditions on the parameters in (35), such that we get a solution

$$\begin{aligned}
\Psi &= \sum_{-\infty}^{\infty} q^n \left( A \sin_q q^{2n+a} + B \cos_q q^{2n+a} \right) \Phi_{2n} \\
&\quad \pm \mathrm{i} \sqrt{q} \sum_{-\infty}^{\infty} q^n \left( B \sin_q q^{2n+1+a} - A \cos_q q^{2n+1+a} \right) \Phi_{2n+1},
\end{aligned} \tag{37}$$

having the eigenvalue $\lambda = \pm \frac{q^a}{\pi_0 (q - 1/q)}$.

Since up to now this is only a formal solution, we have to check the convergence of $\|\Psi\|$ in $H_{\pi_0}$. First, we see that (37) still contains an over-all normalisation constant which we choose to be $A$. With $b = B/A$ we introduce the two functions

$$\begin{aligned}
f_n(a) &= \sin_q q^{2n+a} + b \cos_q q^{2n+a}, \\
g_n(a) &= b \sin_q q^{2n+1+a} - \cos_q q^{2n+1+a}.
\end{aligned} \tag{38}$$

Then the case $A = 0$ is described by $b = \infty$ and appropriate normalisation, which has not to be mixed up with a divergent norm. This means that $b$ has to be considered on the Riemannian sphere, $b \in \hat{\mathbb{C}} = \mathbb{C} \cup \{\infty\}$.

With (38) the formal eigenvector (37) takes the form

$$\Psi = \sum q^n f_n(a) \Phi_{2n} \pm \mathrm{i} \sqrt{q} \sum q^n g_n(a) \Phi_{2n+1}. \tag{39}$$

The norm of this vector is

$$\|\Psi\| = \sum_{-\infty}^{\infty} q^{2n} |f_n(a)|^2 + \sum_{-\infty}^{\infty} q^{2n+1} |g_n(a)|^2. \tag{40}$$

The convergence at $n = -\infty$ of these sums follows directly from the proof of theorem 1. We have to study only $f_n(a)$ and $g_n(a)$ at $n = \infty$. Since $\sin_q z$ and $\cos_q z$ are oscillating functions with increasing period and amplitude, we expect that not for all complex $a, b$ the above norm can be finite. This situation is investigated in the following theorem.

**Theorem 2** *Let $\theta(s)$ the following Theta-function*

$$\theta(s) = \sum_{n=-\infty}^{\infty} (-1)^n q^{-2n^2} q^{ns}. \tag{41}$$

*Then if $b \neq q^{-a} \frac{\theta(2a+1)}{\theta(2a-1)}$ it follows that*

$$f_n(a), g_n(a) \to \infty \qquad \text{for } n \to \infty. \tag{42}$$

*Proof.* Let us consider first

$$f_{2k}(a) = \sin_q q^{4k+a} + b \cos_q q^{4k+a} . \tag{43}$$

Substituting the expansions (33) we get

$$f_{2k}(a) = \sum_{n=0}^{\infty} (-1)^n \frac{q^{-2n^2-n+8nk+2na}}{P_{1,2n+1}} \left( q^{-2n+4k+a-1} + b(1 - q^{-2(2n+1)}) \right), \tag{44}$$

where we denote

$$P_{i,j} = \left(1 - q^{2i}\right)\left(1 - q^{2(i+1)}\right) \cdots \left(1 - q^{2j}\right), \qquad \text{for } i \leq j, \qquad P_{1,0} = 1. \tag{45}$$

Multiplying $f_{2k}(a)$ by

$$s_k = q^{-2(2k+\frac{1}{2}a-\frac{1}{4})^2} P_{1,2k}, \tag{46}$$

we get

$$s_k f_{2k}(a) = \sum_{n=0}^{\infty} (-1)^n \frac{P_{1,2k}}{P_{1,2n+1}} q^{-2(n-2k-\frac{1}{2}a+\frac{1}{4})^2} \left( q^{-2n+4k+a-1} + b(1 - q^{-2(2n+1)}) \right). \tag{47}$$

Let us split this sum in the following parts

$$\sum_{n=0}^{\infty} t_n = \sum_{n=0}^{k-1} t_n + \sum_{n=k}^{\infty} t_n . \tag{48}$$

The first expression tends to zero

$$\left| \sum_{n=0}^{k-1} t_n \right| \leq \sum_{n=0}^{k-1} P_{2n+2,2k} q^{-2(n-2k-\frac{1}{2}a+\frac{1}{4})^2} \left| q^{-2n+4k+a-1} + b(1 - q^{-2(2n+1)}) \right|$$

$$\leq k q^{-2(k+\frac{1}{2}a-\frac{1}{4})^2} \left| q^{4k+a-1} + b \right| \to 0 \quad \text{for } k \to \infty, \tag{49}$$

for all $a, b \in \mathbb{C}$. It is also true for the particular case $b = \infty$, where we have to renormalise. For the second sum we get

$$\sum_{n=k}^{\infty} t_n = \sum_{n=-k}^{\infty} (-1)^n \frac{q^{-2(n-\frac{1}{2}a+\frac{1}{4})^2}}{P_{2k+1,2n+4k+1}} \left( q^{-2n+a-1} + b(1 - q^{-2(2n+4k+1)}) \right). \tag{50}$$

If we let $k \to \infty$ this yields

$$s_k f_{2k}(a) \to \sum_{-\infty}^{\infty} (-1)^n q^{-2(n-\frac{1}{2}a+\frac{1}{4})^2} \left( q^{-2n+a-1} - b \right) \tag{51}$$

$$= q^{-2(\frac{1}{2}a-\frac{1}{4})^2} \left( q^{a-1} \theta(2a-3) + b\theta(2a-1) \right). \tag{52}$$

Hence, $s_k f_{2k}(a) \to 0$ for $k \to \infty$ only if we choose

$$b(a) = -q^{a-1} \frac{\theta(2a-3)}{\theta(2a-1)}, \tag{53}$$

because, as we see later not both $\theta$-functions can be zero for the same $a$. If we examine in the same manner

$$f_{2k+1}(a) = f_{2k}(a+2), \tag{54}$$

we find the condition
$$b(a) = -q^{a+1}\frac{\theta(2a+1)}{\theta(2a+3)}, \tag{55}$$

which is equivalent to (53) because of the quasi-period of the $\theta$-function

$$\theta(s) = \sum_{-\infty}^{\infty}(-1)^n q^{-2n^2} q^{ns} = -q^{-s-2}\sum_{-\infty}^{\infty}(-1)^n q^{-2n^2} q^{n(s+4)} = -q^{-s-2}\theta(s+4). \tag{56}$$

If $b$ does not fullfil the condition (53) we have $s_k f_{2k}(a), s_k f_{2k+1}(a) \to C \neq 0$ and since already $s_k \to 0$ for $k \to \infty$ we can conclude that in this case $f_{2k}(a), f_{2k+1}(a) \to \infty$.
We show finally that the very same applies to $g_k(a)$. From

$$g_k(a) = b\sin_q q^{2n+1+a} - \cos_q q^{2n+1+a} = b\left(\sin_q q^{2n+1+a} + \frac{-1}{b}\cos_q q^{2n+1+a}\right), \tag{57}$$

it is clear that we will get the condition

$$0 = bq^a\theta(2a-1) - \theta(2a+1) \tag{58}$$

or by (56)

$$b = q^{-a}\frac{\theta(2a+1)}{\theta(2a-1)} = q^{-a}\frac{-q^{2a-3+2}\theta(2a-3)}{\theta(2a-1)} = -q^{a-1}\frac{\theta(2a-3)}{\theta(2a-1)}, \tag{59}$$

which is therefore also equivalent to (53) and (55). The first equation here is the condition given in the theorem. $\square$

This theorem tells us that the only possiblity for (39) to have finite norm is to choose

$$b(a) = q^{-a}\frac{\theta(2a+1)}{\theta(2a-1)}. \tag{60}$$

To prove that the norm actually converges in this case we show that any solution of the eigenvalue equation (20) can be casted in the form (39). Then since, as we have shown in theorem 1, there is a unique converging solution, it must be given by (39) with this particular $b(a)$.
In theorem 1 we have shown that $\sum_{-\infty}^{\infty}|c_n|^2 < \infty$ for some particular ratio $c_N/c_{N+1}$ at some large $N$. Here we have

$$\frac{c_{2n}}{c_{2n+1}} = \frac{\pm i}{\sqrt{q}}\frac{\sin_q q^{2n+a} + b\cos_q q^{2n+a}}{\cos_q q^{2n+a+1} - b\sin_q q^{2n+a+1}}. \tag{61}$$

Hence, we have to show that for a fixed $a \in \mathbb{C}$ this expression can take every value in $\hat{\mathbb{C}}$ if we vary $b \in \hat{\mathbb{C}}$. For this purpose we consider (61) as a Moebius transformation in the variable $b$,

$$T(b) = \frac{\alpha + b\beta}{\gamma + b\delta}. \tag{62}$$

Such a linear transformation $T : \hat{\mathbb{C}} \to \hat{\mathbb{C}}$ is bijective, if and only if its determinant is non-zero,

$$\det T = \alpha\delta - \beta\gamma \neq 0. \tag{63}$$

The determinant of the transformation (61) is

$$\det \frac{c_{2n}}{c_{2n+1}} = \frac{\mp i}{\sqrt{q}}\left(\sin_q q^{2n+a}\sin_q q^{2n+a+1} + \cos_q q^{2n+a}\cos_q q^{2n+a+1}\right) = \frac{\mp i}{\sqrt{q}}, \tag{64}$$

which follows from the remarkable identity

$$\sin_q z \sin_q qz + \cos_q z \cos_q qz = 1, \tag{65}$$

proven in the appendix. Therefore the above Moebius transformation is bijective and $\frac{c_{2n}}{c_{2n+1}}$ takes every value in $\hat{\mathbb{C}}$ by varying $b$. Since, by the last theorem, $\sum |c_n|^2 = \infty$ for $b \neq b(a)$ the convergent solution for a fixed $a \in \mathbb{C}$ must be given by $b = b(a)$ in (39).

Let us examine the function $b(a)$ to some more detail. For the $\theta$-function the following product expansion is known [6]

$$\theta(s) = \sum_{-\infty}^{\infty}(-1)^n q^{-2n^2} q^{ns} = (1 - q^{-s-2}) \prod_{n=1}^{\infty}(1 - q^{-4n})(1 - q^{-4n+s+2})(1 - q^{-4n-s-2}). \tag{66}$$

In addition to the quasi-period, $\theta(s) = -q^{-s-2}\theta(s+4)$, this function has the period $\tau = \dfrac{2\pi i}{\log q}$,

$$\theta(s + \tau) = \theta(s), \tag{67}$$

and it is furthermore an even function of the variable $s$,

$$\theta(-s) = \theta(s). \tag{68}$$

Therefore we have also a product expansion for the function $b(a)$,

$$\begin{aligned} b(a) &= q^{-a}\frac{\theta(2a+1)}{\theta(2a-1)} = \\ &= q^{-a}\frac{(1-q^{-2a-3})}{(1-q^{-2a-1})}\prod_{n=1}^{\infty}\frac{(1-q^{-4n+2a+3})(1-q^{-4n-2a-3})}{(1-q^{-4n+2a+1})(1-q^{-4n-2a-1})}. \end{aligned} \tag{69}$$

From the equations (55) and (59) in the above proof we find the property

$$b(a) = -\frac{1}{b(a+1)}, \tag{70}$$

from which it follows that $b(a) = b(a+2)$. Therefore $b(a)$ is an elliptic function with primitive periods 2 and $\tau$,

$$b(a + 2n + \tau m) = b(a), \qquad n, m \in \mathbb{Z}. \tag{71}$$

On the torus $(2, \tau)$ we find for $b(a)$

- zeros  $\quad a = \frac{1}{2}, \quad \frac{1}{2} + \frac{\tau}{2}$
- poles  $\quad a = \frac{3}{2}, \quad \frac{3}{2} + \frac{\tau}{2}$

which can easily read off from the above product expansion. Therefore the function $b(a)$ is of elliptic order 2, which means that it takes any complex values twice on the torus $(2, \tau)$.
Since we need this property for later purpose we calculate these corresponding points. Obviously it is $q^{a+\frac{\tau}{2}} = -q^a$, such that

$$b(a + \frac{\tau}{2}) = -q^a\frac{\theta(2a+\tau+1)}{\theta(2a+\tau-1)} = -b(a), \tag{72}$$

because $\tau$ is a period of $\theta(s)$. On the other hand from $\theta(-s) = \theta(s)$ we find

$$\begin{aligned}
b(-a + \frac{1}{2}) &= q^{a-\frac{1}{2}} \frac{\theta(-2a+2)}{\theta(-2a)} = q^{a-\frac{1}{2}} \frac{\theta(2a-2)}{\theta(2a)} \\
&= -q^{-a-\frac{1}{2}} \frac{\theta(2a+2)}{\theta(2a)} = -b(a + \frac{1}{2}).
\end{aligned} \qquad (73)$$

Combining these two relations (72) and (73) we find

$$b(a) = b(-a - \frac{\tau}{2} + 1), \qquad (74)$$

which is the desired result.

Equipped with these properties of $b(a)$ we can simplify the expression for the eigenvectors of $\xi^*$. With (70) we can rewrite

$$g_n(a) = b(a) \sin_q q^{2n+1+a} - \cos_q q^{2n+1+a} = b(a) f_n(a+1), \qquad (75)$$

and since $b(a+2) = b(a)$ we can consider $f_n(a)$ being a function of the argument $2n + a$,

$$f(2n + a) = f_n(a) = \sin_q q^{2n+a} + b(2n + a) \cos_q q^{2n+a}. \qquad (76)$$

Hence, we have for an eigenvector

$$\Psi(q^a) = \sum q^{\frac{2n+a}{2}} f(2n + a) \Phi_{2n} + ib(a) \sum q^{\frac{2n+1+a}{2}} f(2n + 1 + a) \Phi_{2n+1}. \qquad (77)$$

It has actually the eigenvalue $q^a/\pi_0(q - \frac{1}{q})$, the factor $\pi_0(q - \frac{1}{q})$ is omitted because it is a constant. These are all (unormalised) eigenvectors of the operator $\xi^*$ which we proved to exist in theorem 1.

## 6 Orthogonality and completeness

It is clear that not all eigenstates $\Psi(q^a)$ of $\xi^*$ can be orthogonal in $H_{\pi_0}$. In order to find a complete and orthogonal set we have to study the scalar product

$$\begin{aligned}
(\Psi(q^c), \Psi(q^a)) &= \sum q^{2n + \frac{a+\bar{c}}{2}} \overline{f(2n+c)} f(2n + a) \\
&\quad + b(a)\overline{b(c)} \sum q^{2n+1+\frac{a+\bar{c}}{2}} \overline{f(2n+1+c)} f(2n + 1 + a).
\end{aligned} \qquad (78)$$

Here it is sufficient to study only the first sum for arbitrary $a, c \in \mathbb{C}$. Since the function $f$ has only real coefficients, the complex conjugation in (78) acts only on its argument. We show

**Theorem 3** *Let $a, c \in \mathbb{C}$ with $q^{2a} \neq q^{2\bar{c}}$. Then*

$$\sum_{n=-\infty}^{\infty} q^{2n + \frac{a+\bar{c}}{2}} f(2n + \bar{c}) f(2n + a) = q^{\frac{a+\bar{c}}{2}} \frac{q^a b(\bar{c}) - q^{\bar{c}} b(a)}{q^{2a} - q^{2\bar{c}}}. \qquad (79)$$

*Proof.* From the two difference equations (34) one finds

$$\sin_q z = z^{-2} \left( (q + \frac{1}{q}) \sin_q z - \frac{1}{q} \sin_q(q^2 z) - q \sin_q(q^{-2} z) \right), \qquad (80)$$

and the very same for $\cos_q z$. Since $b(a) = b(a+2)$, this relation is also true for the function $f$

$$f(2n+a) = q^{-2(2n+a)}\left((q+\frac{1}{q})f(2n+a) - \frac{1}{q}f(2n+2+a) - qf(2n-2+a)\right). \tag{81}$$

If we apply this identity to the two functions $f$ in (79), respectively, and subtract the two expressions we get

$$(q^{2a} - q^{2\bar{c}})\sum_{K}^{N} q^{2n+\frac{a+\bar{c}}{2}} f(2n+\bar{c})f(2n+a) =$$

$$\sum_{K}^{N} q^{-2n+\frac{a+\bar{c}}{2}}\left\{\frac{1}{q}f(2n+2+\bar{c})f(2n+a) + qf(2n-2+\bar{c})f(2n+a)\right.$$

$$\left. -qf(2n+\bar{c})f(2n-2+a) - \frac{1}{q}f(2n+\bar{c})f(2n+2+a)\right\}, \tag{82}$$

where the limits $N \to \infty$ and $K \to -\infty$ have to be taken. The above sum is of the type

$$S_K^N = \sum_{K}^{N}\left((b_n - a_n) - (b_{n+1} - a_{n+1})\right), \tag{83}$$

with the notations

$$\begin{aligned} a_n &= q^{-2n+1}f(2n+\bar{c})f(2n-2+a), \\ b_n &= q^{-2n+1}f(2n-2+\bar{c})f(2n+a). \end{aligned} \tag{84}$$

Such a sum can be computed directly

$$\sum_{-\infty}^{\infty}\left((b_n - a_n) - (b_{n+1} - a_{n+1})\right) = \lim_{n\to-\infty}(b_n - a_n) - \lim_{n\to\infty}(b_n - a_n), \tag{85}$$

if both limits exist. Since convergence of the norm implies $f(2n-2+a) \to 0$ for $n \to \infty$, we have

$$\lim_{n\to\infty}(b_n - a_n) = 0. \tag{86}$$

For the limit $n \to -\infty$ in (85) we substitute the definition of $f$

$$\begin{aligned}(b_n - a_n) &= q^{-2n+1}\left(f(2n-2+\bar{c})f(2n+a) - f(2n+\bar{c})f(2n-2+a)\right) \\ &= q^{-2n+1}\left(\sin_q q^{2n-2+\bar{c}} + b(\bar{c})\cos_q q^{2n-2+\bar{c}}\right)\left(\sin_q q^{2n+a} + b(a)\cos_q q^{2n+a}\right) \\ &\quad - \left(\sin_q q^{2n+\bar{c}} + b(\bar{c})\cos_q q^{2n+\bar{c}}\right)\left(\sin_q q^{2n-2+a} + b(a)\cos_q q^{2n-2+a}\right).\end{aligned} \tag{87}$$

Let us examine first the terms containing $\sin_q$-functions

$$q^{-2n+1}\left(\sin_q q^{2n-2+\bar{c}}\sin_q q^{2n+a} - \sin_q q^{2n+\bar{c}}\sin_q q^{2n-2+a}\right) \to 0 \quad\text{for}\quad n \to -\infty, \tag{88}$$

because $\sin_q x \to 0$ and $x^{-1}\sin_q x \to (q - \frac{1}{q})^{-1}$ for $x \to 0$.
The same applies to

$$q^{-2n+1}\left(\cos_q q^{2n-2+\bar{c}}\cos_q q^{2n+a} - \cos_q q^{2n+\bar{c}}\cos_q q^{2n-2+a}\right) \to 0 \quad\text{for}\quad n \to -\infty, \tag{89}$$

which can be verified if we recall the expansion $\cos_q x = 1 - x^2[...]$. For the remaining mixed terms we find finally

$$q^{-2n+1}\Big\{ b(\bar{c})\cos_q q^{2n-2+\bar{c}}\sin_q q^{2n+a} + b(a)\sin_q q^{2n-2+\bar{c}}\cos_q q^{2n+a}$$
$$-b(\bar{c})\cos_q q^{2n+\bar{c}}\sin_q q^{2n-2+a} - b(a)\sin_q q^{2n+\bar{c}}\cos_q q^{2n-2+a}\Big\}$$
$$\longrightarrow \quad q^a b(\bar{c}) - q^{\bar{c}} b(a) \quad \text{for} \quad n \to -\infty, \tag{90}$$

which proves the statement of this theorem. $\square$

By the use of this theorem we find now for the scalar product

$$(\Psi(q^c), \Psi(q^a)) = q^{\frac{a+\bar{c}}{2}}\frac{q^a b(\bar{c}) - q^{\bar{c}} b(a)}{q^{2a} - q^{2\bar{c}}} + b(\bar{c})b(a) q^{\frac{a+\bar{c}}{2}+1}\frac{q^{a+1}b(\bar{c}+1) - q^{\bar{c}+1}b(a+1)}{q^{2a+2} - q^{2\bar{c}+2}}$$
$$= q^{\frac{a+\bar{c}}{2}}\frac{b(\bar{c}) - b(a)}{q^a - q^{\bar{c}}}, \tag{91}$$

for $q^{2a} \neq q^{2\bar{c}}$. Actually, this identity holds for $q^a \neq q^{\bar{c}}$ as well. One has only to check the case $q^a = -q^{\bar{c}}$, which can be computed easily by inserting the difference equtions (34) for $\sin_q$ and $\cos_q$ seperately in the expression (78).

From (91) the importance of the function $b(a)$ becomes clear, every scalar product can be computed by the values of $b(a)$. We use this equation to find orthogonal eigenstates of the operator $\xi^*$. Obviously,

$$(\Psi(q^c), \Psi(q^a)) = 0 \quad \Longleftrightarrow \quad b(\bar{c}) = b(a). \tag{92}$$

Two eigenstates are orthogonal if and only if $b$ takes the same value on their eigenvalues. Since we have already found these corresponding points on the torus, namely $b(a) = b(1 - a + \frac{\tau}{2}) = b(a + 2k)$, we have the following orthogonal eigenstates,

$$\Psi(q^{a+2k}), \quad \Psi(q^{1-\bar{a}+\frac{\tau}{2}+2n}), \qquad k, n \in \mathbb{Z} \tag{93}$$

for an arbitrarily fixed $a$ on the torus $(2, \tau)$. Since we are looking for self-adjoint extensions of $\xi$ we are only interested in real eigenvalues. Hence, we choose a real $a \in [0, 2)$ and find the orthogonal eigenstates

$$\Psi(q^{a+2k}) = \sum_{n=-\infty}^{\infty} q^{n+k+\frac{a}{2}} f(2(n+k)+a)\Phi_{2n}$$
$$+ ib(a)\sum_{n=-\infty}^{\infty} q^{n+k+\frac{a+1}{2}} f(2(n+k)+a+1)\Phi_{2n+1},$$
$$\Psi(-q^{1-a+2k}) = -i\sum_{n=-\infty}^{\infty} q^{n+k+\frac{1-a}{2}} f(2(n+k)+1-a)\Phi_{2n}$$
$$- b(1-a)\sum_{n=-\infty}^{\infty} q^{n+k+\frac{2-a}{2}} f(2(n+k)+2-a)\Phi_{2n+1}, \tag{94}$$

where we recall that $q^{\frac{\tau}{2}} = -1$. For the norm of these states we define

$$\sum_{n=-\infty}^{\infty} q^{2(n+k)+a} f^2(2(n+k)+a) = N^2(a) > 0, \tag{95}$$

where the independance from $k$ follows from the sum over all $n$. $N(a) > 0$ follows from the fact that all summands are positive and that $f(x) > 0$ for small values $x$.

Then we define the squared norms
$$\begin{aligned}
\|\Psi(q^{a+2k})\|^2 &= N^2(a) + b^2(a)N^2(a+1) = M^{-2}(a)\,, \\
\|\Psi(-q^{1-a+2k})\|^2 &= N^2(1-a) + b^2(1-a)N^2(2-a) = M^{-2}(1-a)\,.
\end{aligned} \tag{96}$$

By the normalisation
$$\Psi(q^{a+2k}) \longrightarrow M(a)\Psi(q^{a+2k})\,, \tag{97}$$

we arrive at an orthonormal set of eigenstates. In this manner the poles of $b(a)$ are removed, which came frome the choice $b = B/A$ and neglecting the normalisation constant.

Finally we show that for each $a \in [0,2)$ the set of eigenstates (94) is also a complete set in the Hilbert space $H_{\pi_0}$. This is achieved by constructing the inverse transformation back to the eigenstates $\Phi_n$ of the momentum operator $p$.

**Theorem 4** *The inverse $q$-Fourier transformations are given by*
$$\begin{aligned}
S_1\Phi_{2\ell} &= \sum_{k=-\infty}^{\infty} q^{\ell+k+\frac{a}{2}} f(2(\ell+k)+a)\Psi(q^{a+2k}) \\
&+ i\sum_{k=-\infty}^{\infty} q^{\ell+k+\frac{1-a}{2}} f(2(\ell+k)+1-a)\Psi(-q^{1-a+2k})\,,
\end{aligned} \tag{98}$$

*and*
$$\begin{aligned}
S_2\Phi_{2\ell+1} &= -ib(a)\sum_{k=-\infty}^{\infty} q^{\ell+k+\frac{1+a}{2}} f(2(\ell+k)+1+a)\Psi(q^{a+2k}) \\
&- b(1-a)\sum_{k=-\infty}^{\infty} q^{\ell+k+\frac{2-a}{2}} f(2(\ell+k)+2-a)\Psi(-q^{1-a+2k})\,,
\end{aligned} \tag{99}$$

*with the constants*
$$S_1 = N^2(a) + N^2(1-a)\,, \quad S_2 = b^2(1-a)N^2(-a) + b^2(a)N^2(1+a)\,. \tag{100}$$

*The set of eigenfunctions (94) is therefore a complete basis in $H_{\pi_0}$.*

*Proof.* We show the first identity. Substituting the definitions (94) we get for the r.h.s
$$S_1\Phi_{2\ell} = \sum_n \sum_k c_{n,k}\Phi_n\,, \tag{101}$$

with the coefficients
$$\begin{aligned}
\sum_k c_{2n,k} &= \sum_k q^{2k+\ell+n+a} f(2k+2\ell+a)f(2k+2n+a) \\
&+ \sum_k q^{2k+\ell+n+1-a} f(2k+2\ell+1-a)f(2k+2n+1-a) \\
&= \begin{cases} (N^2(a) + N^2(1-a)) & \text{for } n = \ell\,, \\ q^{n+\ell}\left(\dfrac{b(a)}{q^{2\ell}+q^{2n}} + \dfrac{b(1-a)}{q^{2\ell}+q^{2n}}\right) = 0 & \text{for } n \neq \ell\,, \end{cases}
\end{aligned} \tag{102}$$

which follows from theorem 3 and from $b(a) = -b(1-a)$.
In the same manner we compute

$$\begin{aligned}
\sum_k c_{2n+1,k} &= ib(a) \sum_k q^{2k+\ell+n+a+\frac{1}{2}} f(2k+2\ell+a) f(2k+2n+a+1) \\
&\quad - ib(1-a) \sum_k q^{2k+\ell+n+\frac{3}{2}-a} f(2k+2\ell+1-a) f(2k+2n+2-a) \\
&= ib(a) q^{\ell+n+\frac{1}{2}} \frac{q^{2\ell} b(1+a) - q^{2n+1} b(a)}{q^{4\ell} - q^{4n+2}} \\
&\quad - ib(1-a) q^{\ell+n+\frac{1}{2}} \frac{q^{2\ell} b(-a) - q^{2n+1} b(1-a)}{q^{4\ell} - q^{4n+2}} = 0,
\end{aligned} \qquad (103)$$

where we applied $b(a)b(1+a) = -1$. This proves the first of the equations in the theorem. The second is shown analogously. □

This theorem completes the analytic part of the theory of $q$-deformed Fourier transformations based on the $q$-Heisenberg algebra (1). We summarise our results in the following

**Theorem 5 ($q$-Fourier Transformations.)** *Let $H_{\pi_0} = \{\sum c_n \Phi_n\} : \sum |c_n|^2 \leq \infty\}$ the Hilbert space generated by the orthonormal basis $\{\Phi_n : n \in \mathbb{Z}\}$ of momentum eigenstates, $p\Phi_n = \pi_0 q^n \Phi_n$.*
*Then for each fixed $a \in [0, 2)$ the set of vectors*

$$\begin{aligned}
\Psi(q^{a+2k}) &= \sum_{n=-\infty}^{\infty} q^{n+k+\frac{a}{2}} f(2(n+k)+a) \Phi_{2n} \\
&\quad + ib(a) \sum_{n=-\infty}^{\infty} q^{n+k+\frac{a+1}{2}} f(2(n+k)+a+1) \Phi_{2n+1}, \\
\Psi(-q^{1-a+2k}) &= -i \sum_{n=-\infty}^{\infty} q^{n+k+\frac{1-a}{2}} f(2(n+k)+1-a) \Phi_{2n} \\
&\quad - b(1-a) \sum_{n=-\infty}^{\infty} q^{n+k+\frac{2-a}{2}} f(2(n+k)+2-a) \Phi_{2n+1}, \qquad (104)
\end{aligned}$$

*is also an orthogonal and complete set in $H_{\pi_0}$. They are eigenvectors of the operator $\xi^*$,*

$$\begin{aligned}
\xi^* \Psi(q^{a+2k}) &= \frac{q^{a+2k}}{\pi_0(q - \frac{1}{q})} \Psi(q^{a+2k}), \\
\xi^* \Psi(-q^{1-a+2k}) &= \frac{-q^{1-a+2k}}{\pi_0(q - \frac{1}{q})} \Psi(-q^{1-a+2k}). \qquad (105)
\end{aligned}$$

$$(106)$$

*Together with the inverse transformations in theorem 4 the formulas (104) provide a family of $q$-deformed Fourier transformations in $H_{\pi_0}$ dependent from the real parameter $a \in [0, 2)$.*

We note that the parameter $a$ enters differently the positive and negative eigenvalues of $\xi^*$. In particular, this makes it impossible to rescale the parameter $\pi_0$ by using $a$. Hence, two inequivalent momentum bases characterised by two parameters * $\pi_0$ and $\pi_0'$ have always different families of Fourier transforms.

---
*of the same sign

# 7 Concluding remarks

We first consider the Fourier transformations for the two parameters $a = \frac{1}{2}$ and $a = \frac{3}{2}$, respectively.
For $a = \frac{1}{2}$ we find

$$
\begin{aligned}
f(2n + \frac{1}{2}) &= \sin_q q^{2n+\frac{1}{2}}, \\
b(\frac{1}{2})f(2n+1 + \frac{1}{2}) &= -\cos_q q^{2n+1+\frac{1}{2}},
\end{aligned} \tag{107}
$$

and therefore the eigenstates

$$
\begin{aligned}
\Psi(\pm q^{2k+\frac{1}{2}}) &= \sum_n q^{n+k} \sin(q^{2n+2k}; q^{-4}) \Phi_{2n} \\
&\mp i \sum_n q^{n+k+1} \cos(q^{2n+2k+2}; q^{-4}) \Phi_{2n+1},
\end{aligned} \tag{108}
$$

where we used the notation $\cos(z; q^{-4}) = \cos_q z/\sqrt{q}$ and $\sin(z; q^{-4}) = \sqrt{q}\cos_q \sqrt{q} z$.
For the case $a = \frac{3}{2}$ one has to recall that with $b(\frac{3}{2}) = \infty$, the normalisation constant has a zero of the same order $M(\frac{3}{2}) = 0$. Therefore we find

$$
\begin{aligned}
M(\frac{3}{2})f(2n + \frac{3}{2}) &= \cos_q q^{2n+\frac{3}{2}}, \tag{109} \\
M(\frac{3}{2})b(\frac{3}{2})f(2n+1 + \frac{3}{2}) &= \sin_q q^{2n+1+\frac{3}{2}}, \tag{110}
\end{aligned}
$$

which yields the eigenstates

$$
\begin{aligned}
\Psi(\pm q^{2k-\frac{1}{2}}) &= \sum_n q^{n+k} \cos(q^{2n+2k}; q^{-4}) \Phi_{2n} \\
&\pm i \sum_n q^{n+k+} \sin(q^{2n+2k}; q^{-4}) \Phi_{2n+1}.
\end{aligned} \tag{111}
$$

These two sets of eigenstates (108), (111) have already been found in [2]. There they have been deduced from a $q$-Fourier transformation found by Koornwinder und Swarttouw [4], which is a special case of the Fourier theory developed in this paper.

We can now construct the self-adjoint representations of the operator $\xi$ mentioned in section 4. Defining the vector spaces

$$
E_a = \text{span}\{\Psi(q^{a+2k}), \Psi(-q^{1-a+2\ell}) : k, \ell \in \mathbb{Z}\}, \tag{112}
$$

we get the extensions

$$
\xi_a : D_{\pi_0} \cup E_a \longrightarrow H_{\pi_0}. \tag{113}
$$

where $\xi_a$ on $E_a$ is defined by the corresponding eigenvalues. The operator $\xi_a$ is still symmetric and additionally diagonal on a basis of $H_{\pi_0}$, namely the basis of $E_a$. Hence, as shown in section 3, $\xi_a$ is essentially self-adjoint.
In this way we get for each $a \in [0, 2)$ a different self-adjoint extension of $\xi$ each of which is assigned to a different Fourier transformation $\mathcal{F}_a$. Since clearly $E_{a+2} = E_a$, the parameter $a$ corresponds to a parameter of the group $U(1)$.

The fact that there exists an entire family of Fourier transformations is important if we study the operator $\xi^2$. In the particular cases $a = \frac{1}{2}, \frac{3}{2}$ we find degenerate eigenvalues for $\xi^2$ because, for example, $(q^{\frac{1}{2}+2k})^2 = (-q^{1-\frac{1}{2}+2k})^2$. This is not true in the general case $a \in [0, 2)$.

A problem arises when we study physical theories, where, in general, we have to deal with functions of all variables, for example a Hamiltonian $H = H(\xi, p, u)$. Then one has to require that all operators $p$, $\xi$ and $u$ have to be defined on a common dense domain on which $p$ and $\xi$ are at least essentially self-adjoint. Such an attempt fails here because it is impossible to define $u$ in any of the above extensions of $\xi$. One actually finds that $u$ maps to a different extension, $u : E_a \to E_{a+1}$. Analogously it is not possible to define $p$ on $E_a$.

That something has to be added is already apparent from the different kind of spectra of $p$ and $\xi_a$. We found that the eigenvalues of $\xi_a$ have both sign whereas $p$ has only positive eigenvalues. On the other hand both operators enter the $q$-deformed Heisenberg (1) algebra symmetrically. This problem can be removed if we restart our considerations from the very beginning with a Hilbert space $\mathcal{H}_{\pi_0} = H_{\pi_0} \oplus H_{-\pi_0'}$ with both signs of momentum eigenvalues. Then with the same Fourier transformations on both parts we would end up with extensions $\mathcal{E}_a$ having the property $\mathcal{E}_{a+1} = \mathcal{E}_a$, on which then the operators $u$ and $p$ can be defined. By this procedure the range of $a$ is reduced, $a \in [0, 1)$.

This problem will be discussed in detail in a forthcoming publication.

# 8  Appendix

**1. Worpitzky's Theorem.** *Let $a_2, a_3, a_4, \ldots \in \mathbb{C}$ and*

$$|a_{n+1}| \leq \frac{1}{4}, \qquad \text{for all } n = 1, 2, 3, \ldots \tag{114}$$

*Then the continued fraction*

$$z = \cfrac{1}{1 + \cfrac{a_2}{1 + \cfrac{a_3}{1 + \cfrac{a_4}{\ddots}}}} \tag{115}$$

*converges and its value $z$ is in the domain*

$$\left|z - \frac{4}{3}\right| \leq \frac{2}{3}. \tag{116}$$

A proof of this theorem and a detailed account to the theory of continued fractions can be found in [10].

**2.** To prove the equation $\sin_q z \sin_q qz + \cos_q z \cos_q qz = 1$ we show first a $q$-deformed binomial theorem. We recall the definitions

$$[n]! = (q - q^{-1})(q^2 - q^{-2})(q^2 - q^{-2}) \cdots (q^n - q^{-n}) \tag{117}$$

and $[0]! = 1$. Let us consider the polynomials

$$\begin{aligned} H_0(z) &= 1 \\ H_1(z) &= 1 + z \\ H_2(z) &= (q^{-1} + z)(q + z) \end{aligned}$$

$$H_3(z) = (q^{-2} + z)(1 + z)(q^2 + z)$$

$$\ldots$$

$$H_n(z) = (q^{-(n-1)} + z)(q^{-(n-1)+2} + z) \cdots (q^{(n-1)} + z) \tag{118}$$

Then it holds the following $q$-deformed binomial formula

$$H_n(z) = \sum_{k=0}^{n} \frac{[n]!}{[k]![n-k]!} z^k . \tag{119}$$

*Proof.* From the definition the polynomials $H_n(z)$ one can easily verify

$$(q^{-n} + z) q^n H_n\left(\frac{z}{q}\right) = H_{n+1}(z) , \tag{120}$$

from which the equation (119) follows by straightforward induction. □

Let us consider the following $q$-deformed exponential

$$E_q(z) = \sum_{n=0}^{\infty} \frac{z^n}{[n]!} , \tag{121}$$

for which we find that

$$E_q(\mathrm{i}x) E_q(-\mathrm{i}y) = \sum_{n=0}^{\infty} \frac{(\mathrm{i}x)^n}{[n]!} \sum_{k=0}^{\infty} \frac{(-\mathrm{i}y)^k}{[k]!} = \sum_{n=0}^{\infty} \frac{(-\mathrm{i}y)^n}{[n]!} \sum_{k=0}^{n} \frac{[n]!}{[k]![n-k]!} \left(-\frac{x}{y}\right)^k$$

$$= \sum_{n=0}^{\infty} \frac{(-\mathrm{i}y)^n}{[n]!} H_n\left(-\frac{x}{y}\right) . \tag{122}$$

The exponential decomposes in the trigonometric functions defined in section 4

$$E(\mathrm{i}x) = \cos_q x + \mathrm{i} \sin_q x . \tag{123}$$

such that we get for the product

$$E_q(\mathrm{i}x) E_q(-\mathrm{i}y) = \cos_q x \, \cos_q y + \sin_q x \, \sin_q y + \mathrm{i}(\sin_q x \, \cos_q y - \cos_q x \, \sin_q y) . \tag{124}$$

If we specialise now $y = qx$ and compare then the last equation with (122) we find for the even part

$$\sin_q x \, \sin_q qx + \cos_q x \, \cos_q qx = \sum_{n=0}^{\infty} (-1)^n \frac{y^{2n}}{[2n]!} H_{2n}\left(-\frac{1}{q}\right) = 1 \tag{125}$$

because from the definition (118) of the polynomials, obviously $H_{2n}(-\frac{1}{q}) = 0$ for all $n = 1, 2, 3, \ldots$
The identity (125) assigns next-neighbors on the exponentially spaced lattice.